\documentclass[12pt,preprint]{aastex}
\shorttitle{A New CMD for Pal 11}
\shortauthors{Lewis, Liu \& Chaboyer}

\begin{document}

\title{A New Color-Magnitude Diagram for Palomar 11}

\author{Matthew S. Lewis, W. M. Liu\altaffilmark{1}, N.E.Q.\ Paust 
and Brian Chaboyer}
\affil{Department of Physics and Astronomy, Dartmouth College, 
6127 Wilder Lab, Hanover, NH 03755}

\altaffiltext{1}{present address: Department of Astronomy, University
of Arizona, 933 N Cherry Ave., Tucson AZ 85721-0065}

\begin{abstract}

We present new photometry for the Galactic thick disk globular cluster
Palomar 11 extending well past the main sequence turn-off in the V and
I bands.  This photometry shows noticeable, but depleted red giant
and subgiant branches.  The difference in magnitude between the red
horizontal branch (red clump) and the subgiant branch is used to
determine that Palomar 11 has an age of $10.4\pm 0.5 $ Gyr.  The red
clump is used to derive a distance $\mathrm {d}_\sun=14.3\pm0.4 $ kpc,
and a mean cluster reddening of $\mathrm{ E(V-I)}=0.40\pm0.03$.  There 
is differential reddening across the cluster, of order 
$\delta\, \mathrm{ E(V-I)} \sim 0.07$.  The colour
magnitude diagram of Palomar 11 is virtually identically to that of the thick
disk globular cluster NGC 5927, implying that these two clusters have a
similar age and metallicity.  Palomar 11 has a slightly redder red
giant branch than 47 Tuc, implying that Palomar 11 is 0.15 dex more
metal-rich, or 1 Gyr older than 47 Tuc. Ca II triplet observations 
\citep{RHS97} favour the hypothesis that Palomar 11 is the same age 
as 47 Tuc, but slightly more metal-rich.

\end{abstract}


\section{Introduction}

The globular cluster Palomar 11 (Pal 11) is located at $\rm
\alpha_{2000}=19^h45^m14.4^s $, $\rm \delta_{2000}=-8\arcdeg 0\arcmin
26\arcsec $ ($b = -8 00$; $\ell = 31.81$) in the constellation Aquila.
It sits roughly 13 kpc from the Sun and 8 kpc from the galactic center
with an estimated metallicity of $\mathrm{[Fe/H]}=-0.4$
\citep[][hereafter H96]{H96}.  It belongs to the population of
metal-rich clusters with $R_{GC} \ga 3\,$kpc often associated with the
thick disk.  However, Pal 11 is located well outside the plane of the
disk at $\mathrm{Z}= -3.5\,$kpc \citep{CER93}.

There has been only one published color magnitude diagram (CMD) for
this cluster \citep[][hereafter OBB01]{OBB01}.  OBB01 did note two other
unpublished CMDs from literature abstracts which derived some cluster
properties for Pal 11; \cite{CER93} and \cite{CS84}.  While these
sources derive some cluster parameters (see Table \ref{tbl1} for a list of
published values), none of them report a value for the main sequence
turnoff.  In this paper a new V,V-I CMD for Pal 11 with photometry
reaching well past the main sequence turnoff is presented and used to
determine the age of Pal 11.  There are currently only 6 other thick disk
clusters with age determinations \citep{SW02,DA05}.

In Section 2 the details of the observations and photometry are discussed.
In \S 3 we present the CMD and briefly discuss the blue straggler
 population, \S 4 determines the cluster parameters, such as
distance, reddening, metallicity, and age.  Finally, \S 5
discusses these findings.

\section{Observations \& Photometric Calculations}

Observations of Palomar 11 were obtained over five nights from
September 4 to September 8, 1999 with the 2.4m Hiltner telescope at
the MDM Observatory at Kitt Peak.  All of the images were taken on a
2048x2048 CCD, though the image only covers 1760x1760
pixels to accommodate a 2 inch filter wheel.  The images were taken at
a scale of 0.275 arcsec per pixel for a field of view of $8\arcmin$ 
square.  Table \ref{tbl2} shows a complete list of observations.

\begin{figure}
\epsscale{1.0} 
\plotone{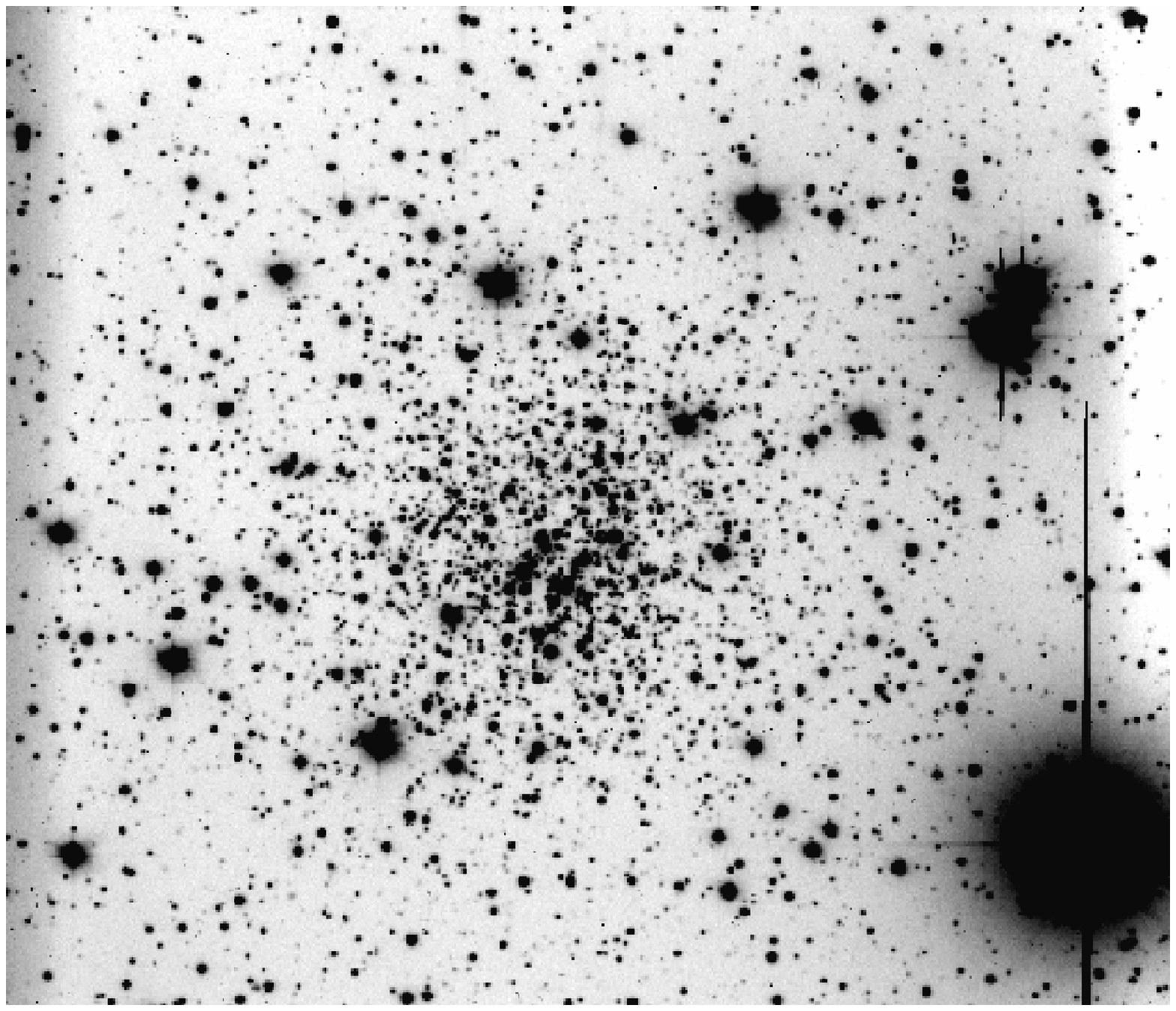}
\caption{Palomar 11 in the V filter.  This image is 484$\arcsec $
square, with North to the right and East up.\label{fig1}}
\end{figure}

\begin{figure}
\epsscale{1.0} 
\plotone{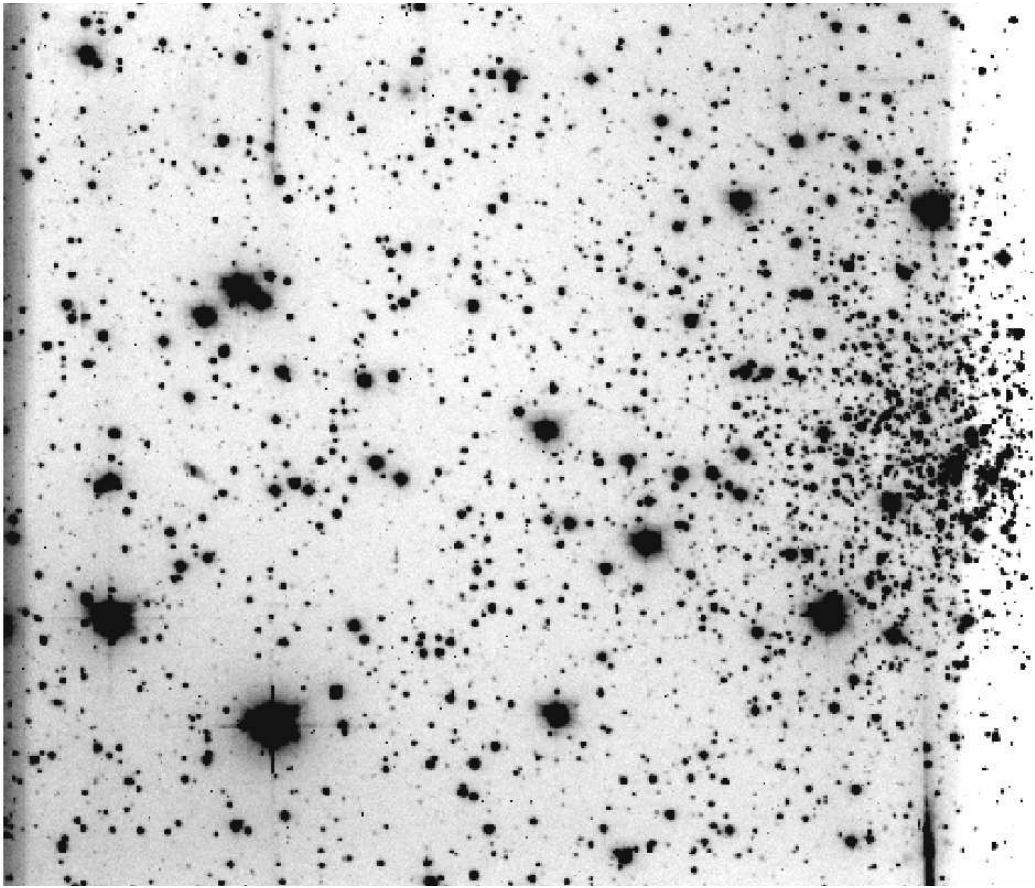}
\caption{The offset field in the V filter.  This image is 484$\arcsec
$ square, with North to the right and East up. \label{fig2}}
\end{figure}

Since the tidal radius of the cluster ($9.8\arcmin$; H96) is
comparable to the image size, observations include images centered on
the cluster as well as images offset a little more than $5\arcmin $ 
South.  This offset field is used to estimate the field star
contamination.  Figures \ref{fig1} and \ref{fig2} are 1200s images in
the V filter of Pal 11 and the offset field.

The image reductions were all done in IRAF\footnote{IRAF (Image
Reduction and Analysis Facility) is distributed by the National
Optical Astronomy Observatory (NOAO), which is operated by the
Association of Universities for Research in Astronomy, Inc., under
contract with the National Science Foundation (NSF).} with its ccdproc
routine.  ALLSTAR and DAOPHOT \citep{ST87,ST94} were used in a
standard manner to do crowded field photometry and extract
instrumental magnitudes. All of images were matched and the
coordinates of all stars were shifted onto a common coordinate system.
To make our final photometry list we require that a star be detected
in a minimum of 2 V and 2 I images.  We find a total number of 6675
stars with V and I magnitudes.  

\begin{figure}
\epsscale{1.0} 
\plotone{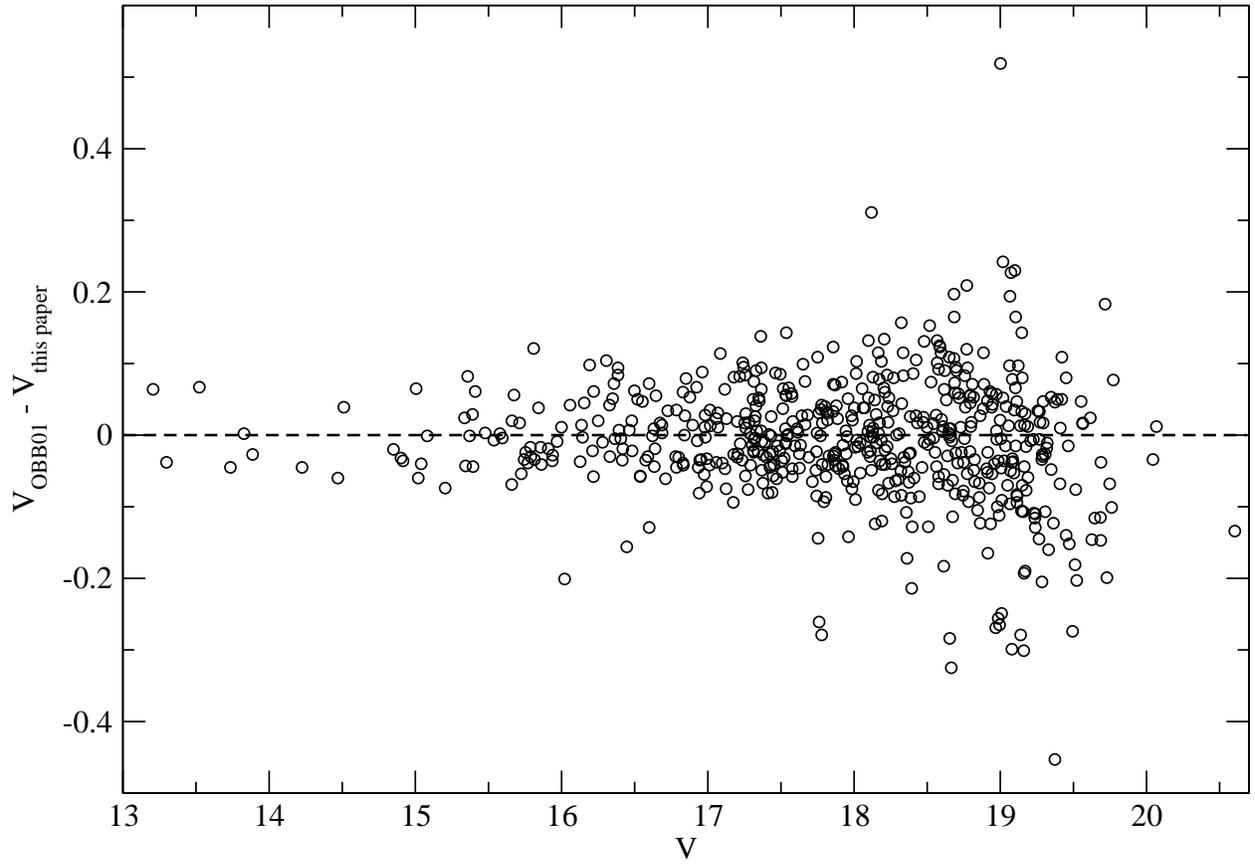}
\caption{The residuals in the matched V photometry between this 
paper and the photometry in OBB01. \label{ortfig}}
\end{figure}

Since none of the nights were photometric, we matched our stars to
those of OBB01 to determine the photometric transformations.  Matches
were found for 583 stars, with $13 < V < 20.6$ and $0.5 < \mathrm{V -
I} < 2.6$.  Our mean photometric errors in the matched stars are 0.024 mag
in V.  Fortunately, there are no apparent trends in the residuals between our
photometry and the OBB01 photometry as a function of magnitude or
colour.  A plot of the magnitude residuals is shown in Figure
\ref{ortfig}.  The bulk of the stars in the match had $16 < V < 19.5$
and $0.9 < \mathrm{V - I} < 1.5$ and it would be difficult to detetect
residual trends with magnitude or colour outside these ranges.  As a
result, our photometric calibration is more uncertain for stars
outside this magnitude and color range.  Fortuantely, the bulk of the
Pal 11 stars fall within this restricted magntiude and colour range.

Our photometry is presented in Table \ref{tbl3}.
The error in the photometry is based on the frame-to-frame variation in
magnitude. The origin of the pixel reference frame is located in the
Southwest corner of the cluster frames.

\section{Color-Magnitude Diagram \& Observational Parameters}

\begin{figure}
\epsscale{0.53}
\plotone{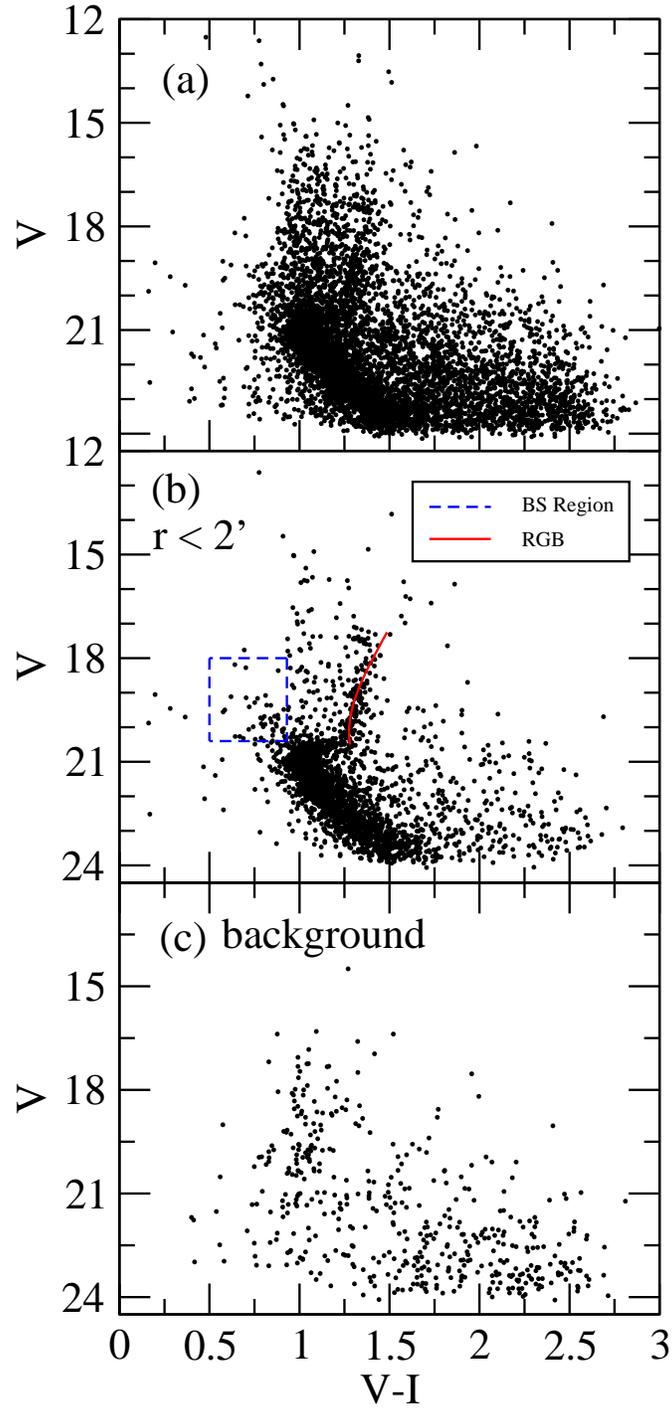}
\caption{Figure (a) shows the CMD for all of the stars in the
field of Palomar 11 and the offset field.  Figure (b) shows only the stars
within 2$\arcmin $ from the center of the cluster and contains the fit
to the RGB and a box defining our BS region. Figure (c) shows the stars 
in an equivalent area to the cluster center field, but located in the 
corners of the image furthest from the cluster center.
\label{fig3}}
\end{figure}

Figure \ref{fig3}a shows the V,V-I CMD of all stars in our
observations, Figure \ref{fig3}b shows the CMD of the stars within the
half-mass radius (2$\arcmin $; H96) of the cluster's center (X=800 \&
Y=820) and Figure \ref{fig3}c shows the CMD of the stars located in
the equivalent area to the cluster center CMD but from regions on our
image which are located farthest from the cluster center
The center of the cluster was found by simply determining
the peak in the star counts in the X and Y directions.
The main sequence, turn-off, and subgiant branch
(SGB) are all clearly visible in both.  The red giant branch (RGB) and
the extremely red horizontal branch (hereafter red clump or RC) are
also visible, but not as well populated.  The fit to the RGB is
overlaid on figure \ref{fig3}b (see section 4.1).

The RC is sloping, which is an indication of differential reddening.
To examine this, we examined the color of the turn-off as a function 
of position for stars in an annulus between 1$\arcmin$ to  
2$\arcmin$ of the cluster center.  The turn-off was found to be reddest 
NorthEast of the cluster center, with a variation in turn-off colour of 
$\mathrm{V - I} = 1.00$ to $1.07$ from the SouthWest to the NorthEast 
of the cluster center.  

The observational parameters of this CMD are presented in Table
\ref{tbl4}.  In this table, $\rm V_{RC} $ and $\rm I_{RC}$ are the
apparent V and I magnitudes for the RC, $\rm (V-I)_g $ is the (V-I)
color of the RGB at the height of the RC, $\rm V_{SGB} $ is the
apparent V magnitude of the SGB, 0.05 dex redder than the main
sequence turnoff \citep{CH96}, and $\rm V_{TO} $ and $\rm (V-I)_{TO} $
are the apparent V magnitude and color of the main sequence turnoff.
These quantities were derived using the mode of the magnitude or
colour distribution for a given quantity.  Given that the cluster is
centrally concentrated and has differential reddening, these modal
values provide an esitimate of the various photometric quantities
at the cluster center.

To estimate the field star contamination in the cluster CMD,
star counts were made within $2\arcmin $ of the cluster center and an
equivalent area in the farthest corners from the cluster center.  This
`background' area starts $5.2\arcmin$ from the cluster center, so
likely contains some cluster stars.  Within $2\arcmin$ of the cluster
center, there are 2670 stars, while there are 457 in the background
area.  This gives a maximum field star contamination of 17\%.  Due to
crowding, the photometry near the cluster center is not as deep or complete.
Restricting the star counts to $\mathrm{V} < 22$, there are 1561 stars
within $2\arcmin$ of the center and 226 stars in the background
region, yielding a maximum field star contamination of 15\%.

The blue straggler population of Pal 11 can be estimated in a similar
manner.  The photometric error in V and (V-I) is used to define the
blue straggler region in the CMD as stars with magnitudes brighter
than $\mathrm{ V_{SGB}} + 3\, \sigma $ and with colors bluer than
$\mathrm{(V-I)_{TO}} - 1\,\sigma$.  These values are
$\mathrm{V_{BS}}=20.4 $ and $\mathrm{(V-I)_{BS}}=0.93$.  This box is
extended to half a magnitude below the RC ($\mathrm{V}=17.99$) and to
$\mathrm{V-I}=0.5 $ in color (see Figure \ref{fig3}b).  In this region
of the CMD, there are 44 stars within $2\arcmin $ of the cluster
center and 13 in the background region.  Thus, about one-third of the
stars in the blue straggler region are field stars, yielding an
estimated number of blue stragglers of $31\pm 7$.  For comparison,
there are 20 RC stars within $2\arcmin$ of the cluster center and 1
star in the region of the RC in the background field.  This yields a
fraction of blue straggler stars of $\mathrm{F_{BSS} \equiv
N_{BS}/N_{RC}} = 1.6\pm 0.8$.  The blue straggler fraction within a
globular cluster is a function of the integrated cluster luminosity
\citep{PIO04}.  Pal 11 has an integrated absolute visual magnitude of
$\mathrm{M_V} = -7.0$, using the integrated visual magnitude of $V_t =
9.8$ from H96 and the distance modulus derived in \S \ref{dist}.
For its luminosity, Pal 11 has a high blue straggler population compared 
to other globuluar clusters \citep{PIO04}.

\section{Cluster Properties}

\subsection{Distance \& Reddening \label{dist}}

The distance and reddening to Pal 11 can be estimated from the
apparent magnitude of the red clump ($\mathrm {V_{RC}}$) using the
method described by \cite{AL02}.  We determine $\mathrm{M^{RC}_V} $
and $\mathrm{M^{RC}_I} $, using the Hipparcos parallax calibration
combined with a correction for the differences in age and metallicity
($\Delta \mathrm{M}_{\lambda}$) between our stars and the Hipparcos
calibration stars \citep{GS01}.  These values depend on metallicity,
so calculations were made at both $\mathrm{[Fe/H]}=-0.7$ and
$\mathrm{[Fe/H]}=-0.4$ to compare with the range of previously
published values.  Above an age of 9 Gyr, $\Delta
\mathrm{M}_{\lambda}$ has a negligible age dependence, so we assume an
age greater than 9 Gyr to determine corrections (see section 4.3 for a
discussion of age).  These calculations yield values of
$\mathrm{M^{RC}_V}=0.62 $; $\mathrm{M^{RC}_I}=-0.28 $ for
$\mathrm{[Fe/H]}=-0.7$ and $\mathrm{M^{RC}_V}=0.75 $;
$\mathrm{M^{RC}_I}=-0.21 $ for $\mathrm{[Fe/H]}=-0.4$.

Assuming a standard extinction law \citep{AL02} the following are
derived: $\mathrm{(m-M)_V}=16.77\pm0.07$,
$\mathrm{(m-M)_I}=16.38\pm0.04$, $\mathrm{A_V}=1.05\pm0.06$,
$\mathrm{A_I}=0.61\pm0.04$, $\mathrm{E(V-I)}=0.40\pm0.03$,
$\mathrm{E(B-V)}=0.31\pm0.03$, and $\rm d_{\sun}=14.3\pm0.4 $ kpc.
These numbers are in general agreement with other numbers published in
the literature, although our reddening is slighter lower and distance
slightly higher than previous estimates (see Table \ref{tbl1}).

\subsection{Metallicity \& Reddening}

The metallicity of Pal 11 has been determined using spectroscopy of
the Ca II triplet \citep{ADZ92}.  The index used is the reduced
equivalent width of the calcium lines for  RGB stars in a
cluster, denoted $W\arcmin $ \citep{RHS97}.  However the current error
bars for Pal 11 are significant, with $W\arcmin =4.77\pm0.22$.  In
terms of metallicity, this gives roughly $\rm [Fe/H]=-0.7\pm0.2 $
based on the updated calibration of $W\arcmin $ by \cite{KI03}.  Two
clusters with a similar Ca II index are 47 Tuc and NGC 5927, with
$W\arcmin =4.53\pm 0.05$, and $W\arcmin =4.85\pm0.06$, respectively.
These two clusters likely bracket Pal 11 in metallicity.

The color of the RGB depends primarly on metallicity and reddening, 
with a smaller dependance on age.  This fact
has been used by \cite{SJ94} to develop a method to
simultaneously determine [Fe/H] and E(V-I), using indicators based on
the RGB color and slope.  This method requires a polynomial fit
to the RGB.  The fit was determined using a two-step process.
First, a rough fiducial was determined by eye.  A standard third-order 
polynomial fit (with no weighting) was then made to
stars which were within 0.10 in V-I of this fidicual and which do not
belong to the RC.  The leads to the following fit to the RGB
\begin{equation}
 V-I = 31.49-4.34*V+0.207*V^2-0.0033*V^3.
\end{equation}
This fit is superimposed on the data in Figure \ref{fig3}b.  

Using this equation and following the procedure outlined by \cite{SJ94}
and the updated calibrations of \cite{SV00} based on
the \cite{CG97} metallicity scale, the
metallicity and reddening of Pal 11 are determined.  For this purpose, the mean
magnitude of the red clump is used as the magnitude of the horizontal
branch.  We derive a metallicity of $\mathrm{[Fe/H]}=-0.76 $ and a
reddening of $\mathrm{E(V-I)}=0.34$.  This implies $\mathrm{E(B-V)}=0.27$.
These same methods were applied to photometry for 47 Tuc \citep{KL98}
and NGC 5927 \citep{FE00}.  For 47 Tuc we derive a metallicity of
$\mathrm{ [Fe/H]}=-0.76 $ and a reddening of $\mathrm{E(V-I)}=-0.02$.
For NGC 5927 we get $\mathrm{[Fe/H]}=-0.77 $ and a reddening of
$\mathrm{E(V-I)}=0.55$.  These reddenings are universally
lower than published values, and so suggests that the derived reddening values
may not be reliable.  It is not clear why the derived reddenings are lower
than previous estimates, but could be related to the fact that the clusters 
studied here are all of relatively high metallicity, and the emperically 
calibration of the colour of the RGB as a function of reddening and metallicity
may not be reliable for these high metallicity clusters.

The derived metallicities are the same for the three clusters.  
However, 47 Tuc was the most
metal rich cluster used to calibrate this method.  The metallicity
derived for NGC 5927 is lower than that inferred from the Ca II index
$\mathrm{[Fe/H]} = -0.64$ \citep{RHS97}; and from Fe measurements 
based upon Fe II lines $\mathrm{[Fe/H]}=-0.67$ \citep{KI03}.  It is 
considerably lower than that determined by \cite{ZW84}, 
 $\mathrm{[Fe/H]}=-0.37$.

Thus, it
appears that one cannot extrapolate the method of \cite{SJ94} past the
metallicity of 47 Tuc.  It is quite possible that Pal 11 is as
metal-rich as NGC 5927.

\subsection{Cluster Age}

Stellar evolution tracks were constructed using stellar evolution code
described by \cite{cha99,cha01} for masses in the range of 0.5
M$_{\odot}$ to 2.0 M$_{\odot}$ in increments of 0.05 M$_{\odot}$.  The
models include the diffusion of helium and heavy elements.  A solar
calibrated model was calculated yielding $Z_{\odot} = 0.02$ and
$Y_{\odot} = 0.275$ (initial abundances). For other metallicities,
${dY}/{dZ}= 1.5$ was assumed (corresponding to a primordial helium
abundance of $Y_{BBN} = 0.245$).  The solar calibrated mixing length
was used for all the models.  The transformation from the theoretical
temperatures and luminosities to observed colors and magnitudes
was completed using the \cite{van03} color tables.

For these isochrones, the magnitude of the SGB, 0.05 magnitudes redder
than the turnoff was used to calculate the difference in magnitude
between the SGB and the semi-empirical $\mathrm{M_V^{RC}} $ determined
above.  This gives us $\Delta \mathrm{V(SGB-RC)} $ as a function of
age at two different metallicities (see Table \ref{tbl5}).  From Table
\ref{tbl5}, it is clear that $\Delta \mathrm{V(SGB-RC)} $ has 
little sensitivity to metallicity over the range $\mathrm{[Fe/H]} =
-0.4$ to $-0.7$ and so can provide a robust estimate of the age of Pal
11 even though its metallicity is not well constrained.  For Pal 11,
$\Delta \mathrm{V(SGB-RC)}=3.15\pm0.04 $, which gives it an estimated
age of $10.4\pm 0.5$ Gyr.  For comparison, 47 Tuc and NGC 5927 were
analyzed in the same manner yielding $\Delta
\mathrm{V(SGB-RC)}=3.16\pm0.04 $ and
$\Delta\mathrm{V(SGB-RC)}=3.15\pm0.04$, corresponding to ages of
$10.6\pm 0.5$ and $10.4\pm 0.5$ Gyr respectively.

\begin{figure}
\epsscale{0.7}
\plotone{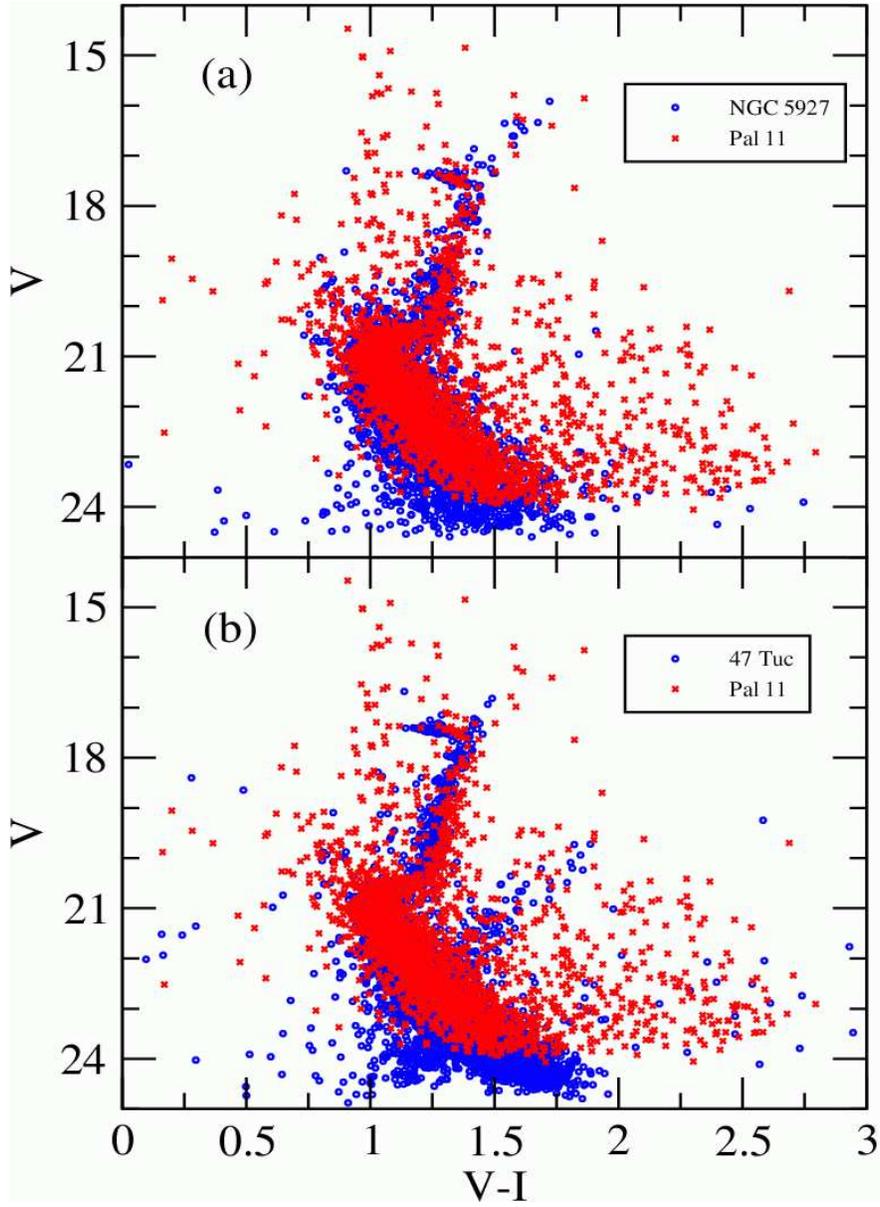}
\caption{Figure (a) shows the CMDs of Pal 11 and NGC 5927, with the
data from NGC 5927 shifted so that the color of the turn-off and magnitude of 
the  SGBs coincide.
Similarly, Figure (b) shows the CMDs of Pal 11 and 47 Tuc.\label{fig4}}
\end{figure}

Figures \ref{fig4}a \& \ref{fig4}b show the CMD of Pal 11 with the
CMDs of 47 Tuc and NGC 5927 overlaid, respectively.  The CMDs of the
two comparison clusters are shifted in color to match the turnoff
color and in V magnitude to match the SGB of Pal 11.  The RGBs and the
RCs of Pal 11 and NGC 5927 overlap, suggesting that they are similar
in age and metallicity.  On the other hand, 47 Tuc has a slightly
bluer RGB suggesting a different age or metallicity (or both).

To estimate the implied age and/or metallicity difference, the
difference in color between the main sequence turnoff and the base of
the RGB as a function of age and metallicity ($\Delta$ (V-I)) was
measured in the isochrones. This index is sensitive to age and
metallicity \citep{SD90,VB90}.  The base of the RGB is defined as the
point where a straight line fit to the RGB crosses the measured value
of $\rm V_{SGB} $ as measured above.  
These theoretical values are compared to the observed data.  Pal 11
has $\Delta\mathrm{(V-I)}=0.17 $ and 47 Tuc has $\Delta
\mathrm{(V-I)}=0.16$, respectively.  Based upon the isochrones, this
difference implies that Pal 11 is 1 Gyr older or 0.15 dex more metal
rich than 47 Tuc.

\section{Summary}

A new CMD for Pal 11 has been presented which reaches roughly 4
magnitudes past the main sequence turnoff. Using the magnitude of the
RC, and the method of \cite{AL02} leads to a distance of
$\mathrm{d}_{\sun}=14.3 \pm 0.4\,$ kpc and a mean reddening of $\mathrm{
E(V-I)}=0.40\pm0.03$. These values are in general agreement with
previously published values. There is differential reddening across cluster, 
such that stars in in the NorthEast of the cluster are approximately
0.07 mag redder in V -- I than stars in the SouthWest part of the cluster.
Following the method of \cite{SJ94}, we
find that Pal 11 has a similar metallicity to the metal rich
clusters 47 Tuc and NGC 5927.

Using the difference in magnitude between the SGB and RC, an age of
$10.4\pm0.5 $ Gyr is determined for Pal 11.  This is the first age
estimate for Pal 11.  Very similar ages are found
for 47 Tuc and NGC 5927.  After shifting to a common reddening and
distance modulus, the CMDs of Pal 11 and NGC 5927 are nearly
identical, which implies that these clusters have the same age and
metallicity.  In contrast, the RGB of 47 Tuc is slighter bluer than
the Pal 11 RGB.  This implies that Pal 11 and 47 Tuc must differ in
age by 1 Gyr or in metallicity by 0.15 dex.  Given that the CMDs of NGC
5927 and Pal 11 are very similar and NGC 5927 has a higher Ca II
triplet equivalent width than 47 Tuc, it is likely that Pal 11 has a
similar age to 47 Tuc, but is slightly more metal-rich.  In agreement
with previous studies \citep{SW02,DA05} we find no evidence for an age
dispersion among the thick disk clusters in the Milky Way.  This
suggests that the stars in the thick disk clusters formed over a relatively
short time-scale ($\la 500\,$Myr).

\acknowledgments

We thank S.\ Ortolani for the electronic version of his Pal 11 photometry
and the anonymous referee whose comments improved the presentation of 
this paper.  Research
supported in part by a NSF CAREER grant 0094231 to BC.  BC is a
Cottrell Scholar of the Research Corporation.

\clearpage

\clearpage

\clearpage

\clearpage

\clearpage

\clearpage

\begin{deluxetable}{crrrr}
\tablewidth{0pt}
\tablecaption{Published Palomar 11 Parameters. \label{tbl1}}
\tablehead{
\colhead{Paper}   &   \colhead{$\rm V_{HB}$}   &   
\colhead{$\rm E(B-V)$}   & \colhead{$\rm [Fe/H]$}   &   
\colhead{$\rm d_\sun$}}
\startdata
OBB01 & 17.4 & 0.35 & -0.7 & 13.2 \\
H96 & 17.40 & 0.35 & -0.39 & 13 \\
CER93 & 17.35 & - & -0.23 & - \\
ADZ92 & - & -  & $>$-0.6 & - \\
W85 & 17.38 & 0.34 & -0.92\tablenotemark{\dagger} & 13.8 \\
Z85 & & 0.35 & -0.7 & - \\
CS84 & 17.3 & 0.34 & - & 12.9 \\
\enddata
\tablenotetext{\dagger}{This metallicity is [M/H] rather than [Fe/H]}
\tablecomments{(\citealt{OBB01}; \citealt{H96} 
from the February 2003 on-line edition; \citealt{
CER93,ADZ92,W85,Z85,CS84})}

\end{deluxetable}

\clearpage

\begin{deluxetable}{cccrrr}
\tablewidth{0pt}
\tablecaption{Log of Observations. \label{tbl2}}
\tablehead{
\colhead{Target}         & \colhead{Date}  &
\colhead{Filter}         & \colhead{Time(s)}  &
\colhead{Number}         & \colhead{FWHM(\arcsec)}}
\startdata
Pal 11 & 9/4/99 & I & 20 & 3 & 1.2\\
 & `` & I & 120 & 3 & 1.1\\
 & `` & I & 900 & 2 & 1.4\\
 & `` & V & 20 & 3 & 1.1\\
 & `` & V & 200 & 3 & 1.0\\
 & `` & V & 1200 & 3 & 1.4\\
 & 9/5/99 & I & 15 & 1 & 0.8\\
 & `` & I & 900 & 4 & 1.3\\
 & `` & V & 1200 & 1 & 1.4\\
 & `` & V & 229 & 1 & 1.3\\
 & 9/6/99 & I & 20 & 1 & 1.2\\
 & `` & V & 20 & 1 & 1.3\\
 & 9/7/99 & V & 20 & 1 & 1.2\\
 & `` & V & 1200 & 2 & 1.3\\
Offset & 9/6/99 & I & 20 & 4 & 1.3\\
 & `` & I & 120 & 4 & 1.3\\
 & `` & I & 900 & 3 & 1.3\\
 & `` & V & 20 & 3 & 1.2\\
 & `` & V & 200 & 4 & 1.4\\
 & `` & V & 1200 & 2 & 1.2\\
 & 9/7/99 & I & 900 & 3 & 1.0\\
 & `` & V & 1200 & 3 & 1.3\\
\enddata


\end{deluxetable}

\clearpage

\begin{deluxetable}{rrrrrrr}
\tablewidth{0pt}
\tablecaption{Photometry of Palomar 11 \label{tbl3}}
\tablehead{
\colhead{Number} & \colhead{X pixel}    &  \colhead{Y pixel}  & 
\colhead{V}      & \colhead{$\delta\,$V\tablenotemark{\ast}}   
& \colhead{I}  & \colhead{$\delta\,$I\tablenotemark{\ast}}}
\startdata
1 & -617.12 & -49.09 & 18.687 & 0.018 & 17.510 & 0.028\\
2 & -239.37 & -47.76 & 22.975 & 0.079 & 20.921 & 0.052\\
3 & -450.70 & -44.58 & 20.349 & 0.018 & 19.267 & 0.016\\
4 & -760.36 & -43.60 & 19.955 & 0.011 & 18.937 & 0.015\\
5 & 704.18 & -42.28 & 22.573 & 0.051 & 20.493 & 0.032\\
\enddata
\tablenotetext{\ast}{These errors are determined by the frame to frame
variation in V or I}
\tablecomments{A full table of photometry can be found in the
electronic version of this article}
\end{deluxetable}

\clearpage

\begin{deluxetable}{cr}
\tablewidth{0pt}
\tablecaption{Palomar 11 Parameters. \label{tbl4}}
\tablehead{\colhead{Parameter}  &  \colhead{Value}}
\startdata
$\rm V_{RC}$  & $17.46\pm 0.03$\\
$\rm I_{RC}$  & $16.14\pm 0.03$\\
$\rm (V-I)_{g}$  & $1.43\pm 0.02$\\
$\rm V_{SGB}$  & $20.61\pm 0.02$\\
$\rm V_{TO}$  & $20.88\pm 0.06$\\
$\rm (V-I)_{TO}$  & $1.03\pm 0.01$\\
\enddata
\end{deluxetable}

\clearpage

\begin{deluxetable}{ccc}
\tablewidth{0pt}
\tablecaption{Theoretical $\Delta$V(SGB-RC)\label{tbl5}}
\tablehead{
\colhead{Age}    &  \multicolumn{2}{c}{$\Delta$V(SGB-RC)}\\
\colhead{(Gyr)} & \colhead{[Fe/H]=$-$0.4}   & \colhead{[Fe/H]=$-$0.7}}
\startdata
~9  &  3.002  &  2.989\\
10  &  3.114  &  3.104\\
11  &  3.218  &  3.189\\
12  &  3.304  &  3.278\\
13  &  3.385  &  3.358\\
14  &  3.461  &  3.443\\
15  &  3.532  &  3.508\\
\enddata
\end{deluxetable}

\end{document}